\documentclass[aps,prl,twocolumn,superscriptaddress,showpacs]{revtex4-1}
\usepackage{hyperref}
\usepackage{graphicx}
\usepackage{algorithm2e}
\usepackage{extarrows}
\usepackage{times}
\usepackage{mathrsfs}
\usepackage[english]{babel}
\usepackage{tabularx}
\usepackage[margin=2cm]{geometry}

\begin{document}

\title{From quantum to classical dipole plasmon resonances in highly-doped nano-crystals}
\author{Leonid Gerchikov}
\affiliation{St.Petersburg National Research Academic University RAS, 194021 St.Petersburg, Russia}
\author{Andrey Ipatov}
\affiliation{St.Petersburg National Research Academic University RAS, 194021 St.Petersburg, Russia}
\author{Claude Guet}
\email[Corresponding author: ]{cguet@ntu.edu.sg}
\affiliation{Energy Research Institute (ERI@N) and School of Materials Science and Engineering, Nanyang Technological University, 639798 Singapore}
\date{\today }

\begin{abstract}
   Dipole plasmon resonances are ubiquitous in nano-particles with delocalized charge carriers. Doped semi-conductor colloidal nano-crystals constitute a novel paradigm for plasmon excitations in a finite electron system and offer the possibility to tune the carrier density and thus the dipole resonance from visible to infra-red, which cannot be achieved with metallic clusters. Restricting ourselves to highly n-doped ZnO nano-crystals, we explain the observed smooth transition from small sizes dominated by quantum effects to large sizes where the resonance reaches its classical value. A schematic two interacting highly degenerate level quantum model, validated by a full Random Phase Approximation calculation, yields nicely the experimentally observed trends.
\end{abstract}
\maketitle

Dipole plasmon excitations in metallic nano-particles occur at visible and UV frequencies as the small oscillation of the delocalized valence electrons implies carrier densities of typically $10^{22} \div 10^{23}$ cm$^{-3}$. For simple metals, the spill-out of electrons leads to a red shift of the resonance frequency as the particle size decreases, with additional quantum fluctuations due to electronic shell effects observed only for particles with less than a few hundred of atoms~\cite{Brack}. For noble metals, the induced d-electrons polarization cannot be neglected as it modifies quantitatively the above picture, but not qualitatively. Phenomenologically, the asymptotic dipole plasmon frequency can be estimated within the Drude model and taking the empirical dielectric function of the metal~\cite{Lerme}.

Motivated by plasmonics applications to numerous fields, tuning plasmon frequencies down to the IR range has been the object of intense works over the last ten years, as witnessed by the recent review by Kriegel et al~\cite{Kriegel}. We focus this Letter on "Charge-Tunable Quantum Plasmons in Colloidal Semiconductor Nanocrystals", quoting the title of Ref.~\cite{Alina1}. These authors showed that it was possible to generate highly charged semiconductor ZnO  nano-crystals (NC) by combining photo doping and hole scavenging~\cite{Alina1,Alina2}. They found that the carrier concentration reached some upper limit which was controlled mainly by the hole quencher used. Absorbance was measured over a range from 0.2 to 1 eV, for almost spherical ZnO NC's of radius from about 1 to 6 nm with electron concentrations ranging from about 0.2 to 2 $\times 10^{20}$  cm$^{-3}$. Schimpf et al showed that the measured resonance frequencies diverged from the classical Drude model predictions, whereas a forced damped oscillator with input transition energies and strengths given by a quantum \textbf{non interacting} particle model yielded the correct size dependence only qualitatively~\cite{Alina1}. Another study also emphasized the collective nature of SPR in photocharged ZnO NC's~\cite{Faucheaux1} . The two-fold objective of this Letter is to: $i)$ revisit the theoretical approach and show that it is mandatory to go beyond the non-interacting picture; $ii)$ emphasize the distinctiveness of colloidal highly doped nano-crystals as systems of N interacting electrons inside a Faraday cage.

The paper is organized as follows. A mean field Local Density Approximation (LDA) yields the ground state single-particles. As the radial components of these eigenfunctions are very close to each other and the eigenvalues obey an angular-like spectrum, the dipole response is worked out within a two degenerate interacting level model. This model predicts a collective state carrying two third of the total oscillator strength and of frequency reaching the classical value in the large size limit. It shows that the transition from non-interacting to interacting behavior occurs at a length scale of the order of the effective ZnO Bohr radius. The predictions from this model agree nicely with the measured data of Ref.~\cite{Alina1,Alina2} and are validated by a full Random Phase Approximation (RPA) calculation.

We consider carrier-doped ZnO NC's embedded in toluene as discussed by Schimpf et al~\cite{Alina1}. The electron concentration used in subsequent calculations, $n_e$ = $3N/4 \pi R^3$ = $1.4 \times 10^{20}$ cm$^{-3}$, corresponds to an effective Wigner-Seitz radius $r_s$ = $(3/4\pi n_e)^{1/3}$ = $1.2$ nm. The NC radius, $R$ ranges from 2.4 to 6 nm, accordingly the number of conduction electrons, $N$ from about 10 to 130. It is known that the electronic band structure of bulk ZnO is characterized by a non isotropic and non parabolic energy dispersion\cite{ZnO}. However, for the current problem with many delocalized electrons, we assume that these electrons follow an isotropic parabolic energy dispersion and carry an average effective mass $m_e^*=0.3~m_e$\cite{ZnO}. For the same reason, the envelop function approximation is justified. As the electrons are strongly confined within the ZnO NC by the conduction band offset at the interface~\cite{Monreal}, we can impose the electron wave functions to vanish at the interface and view the colloidal NC as a system of $N$ interacting electrons inside an infinite spherical well, $V_{ext}$, of radius $R$. Note that the overall charge neutrality in ensured by a positive charge surface distribution that does not create any field inside the NC.  The Hamiltonian is simply:
\begin{equation}
\hat{H} = \sum_a \frac{\hat{{\bf p}}^2_a}{2 m_e^*}+
\frac{1}{2}\sum_{a,b}V\left({\bf r}_a,{\bf r}_b \right) + \sum_a V_{ext}(r_a).
\label{H12}
\end{equation}
The Coulomb interaction between an electron at position ${\bf r}_a$ and one at position ${\bf r}_b$ is screened by the polarization of both ZnO material and embedding medium, so that the radial coefficients of its Legendre polynomials expansion
$V\left({\bf r}_a,{\bf r}_b \right)=\sum_LV_LP_L\left({\bf r}_a{\bf r}_b/r_ar_b \right)$
write as:
\begin{equation}
V_L=
\frac{e^2}{\varepsilon_i}\left(\frac{r^L_<}{ r^{L+1}_>}+
\frac{\left(\varepsilon_i-\varepsilon_m\right)\left(L+1\right)\left(r_a r_b\right)^L}{\left(L\varepsilon_i+(L+1)\varepsilon_m\right) R^{2L+1}} \right) ,
\label{V12}
\end{equation}
where $r_<$ and $r_>$ are the smallest and largest of the two radial positions. ZnO and toluene dielectric constants are assigned their bulk values $\varepsilon_i=3.7$ and $\varepsilon_m=2.25$\cite{Alina1}, respectively. With these parameters, the effective Bohr radius $a_0=\hbar^2 \varepsilon_i/m_e^* e^2 = 0.65$ nm, is smaller than the NC radius.

Each electron satisfies the LDA Kohn-Sham quation
\begin{equation}
[\frac {{\bf{\hat{p}}}^2}{2 m_e^*}+ V_{mf}]~ u_a({\bf r}) = \epsilon_{a}  u_a({\bf r}),
\label{hf}
\end{equation}
where $V_{mf}$ is the sum of the direct Coulomb potential, $V_D({\bf r})=\int V({\bf r},{\bf r}^{\prime})\rho({\bf r}^{\prime})d{\bf r}^{\prime}$ and the exchange potential $V_x= -(e^2/\varepsilon_i)(3\rho({\bf r})/\pi)^{1/3}$ with $\rho=\sum_a |u_a|^2$ being the electron density.
\begin{figure}[ht]
\begin{center}
\includegraphics[scale=0.28]{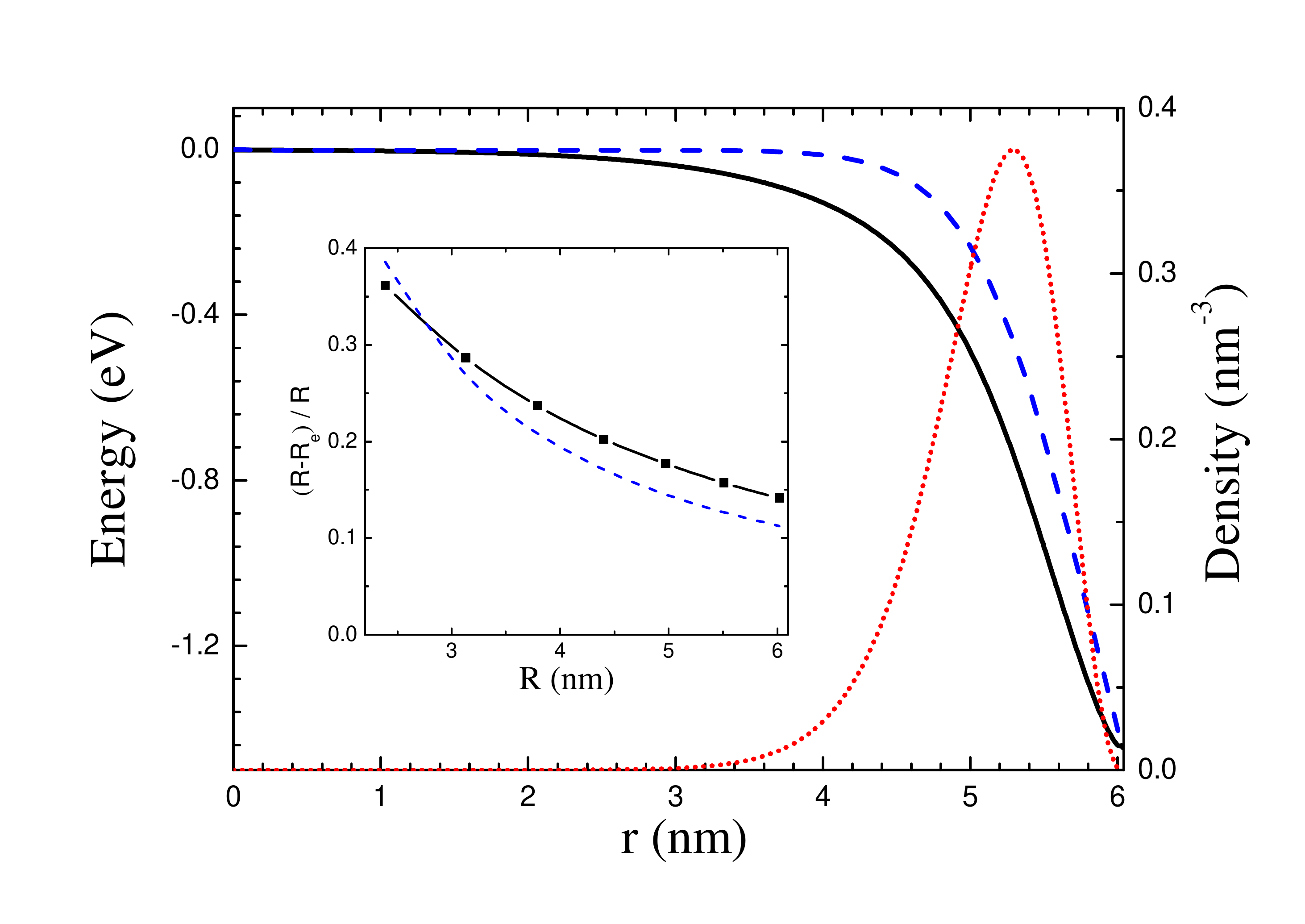}
\end{center}
\caption{Radial profile of a spherical ZnO NC doped with 128 electrons; mean field potential: black line; direct Coulomb potential: blue dash line; density: red dot-dot line. In the insert: Relative position of mean electron radius $R_e$; mean field potential prediction: black line; triangle model, Eq.~(\ref{ReAiry}): blue dash line}
\label{fig1}
\end{figure}
For spherical systems with electrons filling closed shells, the index $a=(n,l)$, where $n$ is the radial quantum number and $l$ the angular momentum one. Fig.~\ref{fig1} shows that the radial density distribution is narrow and shifts towards the edge as the number of electrons increases, for a given $r_s$ parameter. Therefore, the Coulomb repulsion leads to a hollow electron spherical distribution rather than a spherical uniform distribution typical of metallic nano-particles. Whereas the electron distribution "spills out" at the surface of the latter, it is "pushed-in" for the the former, because of quantum pressure. In order to estimate how the electron mean radius $R_e$ varies with $R$ we approximate $V_{mf}$ by a triangle well with a slope equal to the electric field $F=eN/\varepsilon_iR^2$ at the interface. Such an approximation is justified close to the edge, as can be seen from Fig.~\ref{fig1} which shows the radial profile of $V_{mf}$. The 1s solution of the Schr\"{o}dinger equation is expressed in terms of the Airy function\cite{Landau} that vanishes at the boundary, $Ai\left((R-r)/r_0-x_1\right), x_1\simeq 2.34$. This simple model yields an approximate mean radius:
\begin{equation}
R_e^{app} \simeq R - 1.5 r_0 = R - 1.5 r_s(a_0/2R)^{1/3}
\label{ReAiry}
\end{equation}
 where $r_0=r_s(a_0/2R)^{1/3}$ is the length scale of our problem which is much smaller than the NC radius. As shown in the insert of Fig.~\ref{fig1}, this approximate relation between the mean electron radius and the NC radius is close to its mean field value obtained from solving the set of Eqs.~(\ref{hf}).

The large difference between $r_0$ and $R$ has dramatic consequences. First the angular motion does not change the radial density distribution and second  the angular kinetic energy $\hbar^2l(l+1)/2m_e^*R_e^2$ is much smaller than the energy scale of radial motion $E_0 = \hbar^2/2m_e^*r_0^2$. Therefore, the single-particle states with one radial node or more are not populated in the ground state.  This was checked numerically for systems of N electrons, N taking values between 8 to 128 such that all possible closed shell configurations were tested. Unsurprisingly, the lowest energy shell sequence was found to be $1s^2,1p^6,1d^{10}...1l_{max}^{2(l_{max}+1)}$, the radial wave functions of these occupied ''node-less'' states being very similar. It happens because the centrifugal force for all these states is much smaller than the Coulomb repulsion force $eF$, the ratio being $\hbar^2l_{max}(l_{max}+1)/2m_e^*R_e^3/eF \simeq a_0/2R \ll 1$. That is why the radial and angular electron motions separate and the occupied single-particle energy spectrum $\epsilon_{a}$ of Eq.~(\ref{hf}) is well approximated by
\begin{equation}
\epsilon_{1,l} =\epsilon_{1s} +\frac{\hbar^2l(l+1)}{2m_e^* R_e^2}.
\label{rotator}
\end{equation}This electronic structure gives "magic numbers" of electrons $N=2(l_{max}+1)^2$, where $l_{max}$ is the angular momentum of the highest occupied orbital (HOO). We checked that the first excited levels were also radially node-less. Thus, a dipole excitation of the non-interacting system described by Eq.~(\ref{rotator}) creates a hole in the HOO and a particle in the lowest unoccupied orbital (LUO) of angular momentum $l_p=l_{max}+1$, the non-interacting transition energy being
\begin{equation}
\Delta = \left(\epsilon_p -\epsilon_h\right)= \hbar^2 (l_h+1)/m_e^*R_e^2.
\label{Delta}
\end{equation}

Let us apply to the N-electron system a weak harmonic dipole electric field of frequency $\omega$ and seek a solution of the time-dependent one-particle mean-field equation, of the form
\begin{equation}
\psi_a({\bf r},t) = [u_a\left({\bf r}\right)
+ w_a^{+} \left({\bf r}\right) e^{-i \omega t}
+ w_a^{-} \left({\bf r}\right) e^{i \omega t}
] e^{-i \frac{\epsilon _a t }{\hbar}},
\label{tdhf1}
\end{equation}
where $w_a^{\pm}$ are small perturbations. Within our schematic model based on Eq.~(\ref{rotator}),
dipole excitations can only be between the two degenerate HOO $|$$h$$>$ and LUO $|$$p$$>$ states. Discarding the driving terms, we obtain
\begin{equation}
[\frac {{\bf{\hat{p}}}^2}{2 m_e^*} +V_{mf}- (\epsilon_h \pm \hbar\omega)]w_{h^{\pm}}({\bf r}) +  V_{mf^{\pm}}^{(1)} u_h({\bf r}) = 0,
\label{tdhf2}
\end{equation}
where we explicitly keep the first order modifications $V_{mf^{\pm}}^{(1)}$ of the mean field potential and thus adopt the random-phase approximation (RPA)~\cite{Dalgarno,Amusia,Bertsch}. As only the $|$$h$$>$ orbital is perturbed under the action of a dipole field, the first order perturbation is
 \begin{equation}
V_{mf^{\pm}}^{(1)} = \int d{\bf r'}v\left(\bf r,\bf r'\right)\left(u_h^{*} w_h^{\pm} +u_h w_h^{\mp^{*}} \right),
\label{Vmf1}
\end{equation}
where $v\left(\bf r,\bf r'\right)=V\left(\bf r,\bf r'\right)+
 \delta\left(\bf r-\bf r'\right)\delta V_x/\delta \rho$.
 As in the same conditions, only the virtual LUO, $|$$p$$>$, can be excited, the perturbed orbitals reduce to
\begin{equation}
w_{h^+}({\bf r})= X u_p({\bf r}) \quad \textrm{and} \quad   w_{h^-}({\bf r})= Y u_p({\bf r}),
\label{XY}
\end{equation}
where $X$ and $Y$ are RPA forward and backward amplitudes, respectively.
By inserting Eqs.~(\ref{XY}), (\ref{Vmf1}) into Eq.~(\ref{tdhf2}), and taking the scalar product with the bra $<$$p$$|$ we obtain the set of two equations:
\begin{eqnarray}
\nonumber
\left(\Delta + V_A -\hbar\omega\right) X + V_B Y =0 \\
\left(\Delta + V_A +\hbar\omega\right) Y + V_B X =0
\label{2by2}
\end{eqnarray}
where $V_A$$ =$$\langle ph|v|hp \rangle$ is the matrix element between $p$-$h$ excitations and $V_B$$ = $$\langle pp|v|hh \rangle$ is the matrix element of that interaction between the ground state and two-particle-two-hole excitations. This schematic model with only local exchange, $V_A$$ =$$ V_B=V$, yields:
\begin{equation}
\omega^2 = \left(\Delta^2 + 2 \Delta V\right)/ \hbar^2,
\label{omega}
\end{equation}
and shows that the repulsive $p$-$h$ interaction blue shifts the non-interacting transition.
\begin{figure}[ht]
\begin{center}
\includegraphics[scale=0.28]{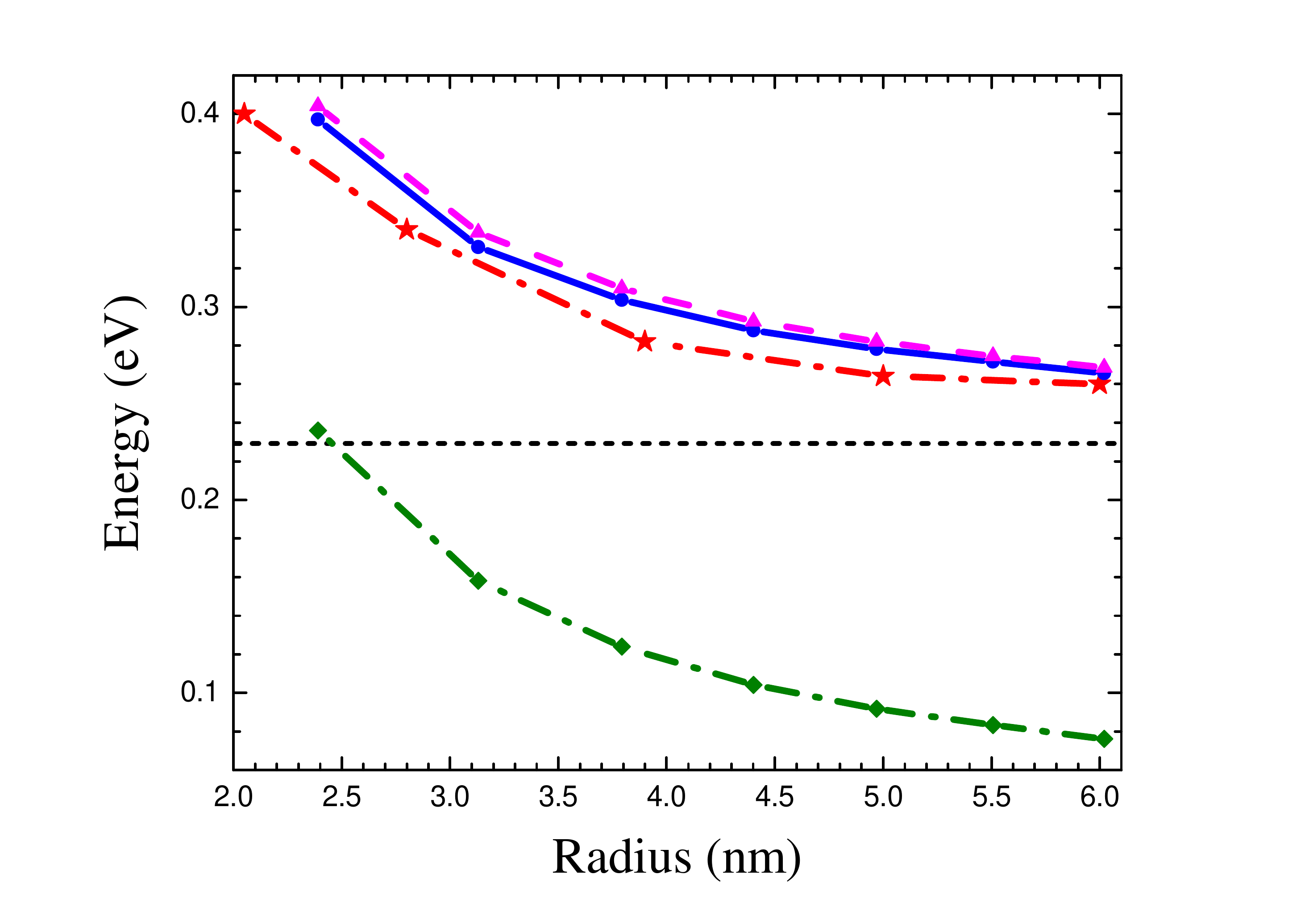}
\end{center}
\caption{Size dependence of the dipole resonance peak frequency. Experimental values \cite{Alina1}: Red stars on dot-dashed line . Two-level model, Eq.~(\ref{omega}): Magenta triangles on dashed line. Full RPA with local exchange: blue circles on full line. Non-interacting model, Eq.~(\ref{Delta}): green diamonds on lower dot-dashed line. Classical plasmon energy, Eq.~(\ref{class}) : horizontal line.}
\label{fig2}
\end{figure}

In the classical limit of a large NC, $R_e\rightarrow R$,
the exchange term in matrix element $V$ can be neglected and the radial part of $V$ is the dipole term in the expansion~(\ref{V12}) with $r_a$$=$$r_b$$=R$, thus $V_{R}= 2e^2(l_h+1)/(\varepsilon_i+2\varepsilon_m)R$ (the factor 2 being due to the sum over spin states) and $\Delta_{R} = (\hbar^2/m_e^*R^2 )\sqrt{N/2}$. Within this limit, the second term of Eq.~(\ref{omega}) writes as:
\begin{equation}
2\Delta_R V_R / \hbar^2 = \frac{2 e^2 N }{m_e^* \left(\varepsilon_i + 2 \varepsilon_m\right)  R^3} =\omega_{cl}^2.
\label{class}
\end{equation}
where $\omega_{cl}$ is nothing but the classical frequency~\cite{Class}, equal to $0.23$ eV/$\hbar$ for ZnO NC's in toluene. Note by passing that the classical squared frequency for the infinitely thin layer is two-third of the squared Mie frequency of a uniform spherical electron distribution of same $r_s$ parameter. In the former case, the collective motion is associated with angular motion within the thin shell, while it is a collective displacement of all electrons along one coordinate axis in the second case. The transition from pure quantum non-interacting mean-field single-particle behaviour to interacting behaviour is thus controlled by the ratio $\lambda$
\begin{equation}
\lambda = \frac{\Delta_R}{2V_R}  = \frac{a_0}{4R}
\left(1 + \frac{2 \varepsilon_m}{\varepsilon_i}\right) \simeq \frac{a_0}{2R},
\label{lambda}
\end{equation}
which turns out to be of the order of $1/10$  in favour of a strongly interacting behaviour as expected from the small ratio of the centrifugal force to the Coulomb repulsion force.
By use of Eqs.~(\ref{2by2}) we note that a small ratio $\lambda$ implies that the backward amplitude $Y$ is as large as $X$ which expresses that the ground state is strongly corrected by the RPA correlation, $V$. Nonetheless, the situation may be different for doped NC's made of materials with a much larger dielectric function and smaller effective mass.

The finite size also affects the dipole resonance frequency. Due to the quantum "push-in" effect, the electron distribution moves away from the surface as the size decreases, see Fig.~\ref{fig1}. The predicted frequencies for finite sizes calculated with Eq.~(\ref{omega}) agree surprisingly well with those measured by Schimpf et al \cite{Alina1} as can be seen in Fig.~\ref{fig2}. The "push-in" effect acting on both the $\Delta$ and $V$ terms explains the pronounced blue shift with decreasing size seen in the experiment.

The strength is a good measure of the collectivity of the excitation. The oscillator strength $f_1$ for the transition from the many-body ground state $|$$0$$\rangle$ of zero total angular momentum to the unique many-body dipole state $|$$1$$\rangle$ of excitation energy $\hbar \omega$, can be expressed in terms of the the single $p$$-$$h$ dipole matrix element as:
 \begin{equation}
f_1 = \frac{2m_e^* \omega}{\hbar}|\langle 1|Z|0 \rangle|^2 = \frac{2m_e^* \Delta}{\hbar^2}|\langle p|z|h \rangle|^2 \simeq \frac{2}{3} N,
\label{f1}
\end{equation}
where $Z=\sum_{i=1}^N z_i$.

As there is no radial kinetic term in two-level model based on Eq.~(\ref{rotator}), the evaluation of the f-sum rule is restricted to two out of three momentum degrees of freedom. Thus, in this model the sum rule should be equal to $2N/3$.
\begin{figure}[ht]
\begin{center}
\includegraphics[scale=0.28]{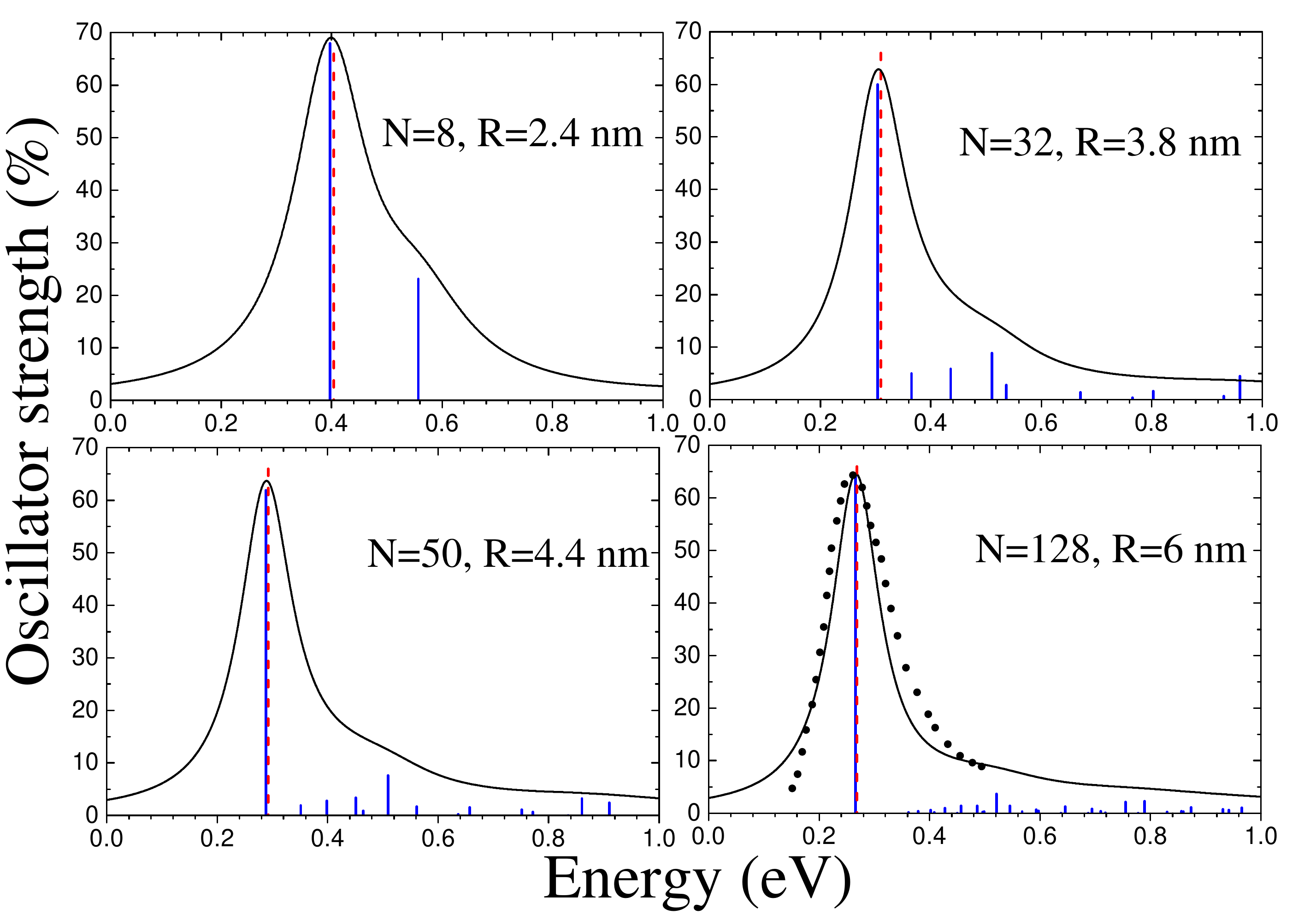}
\end{center}
\caption{Oscillator strength distribution for model closed shell ZnO NC's, calculated in the RPA with local exchange: blue vertical bars. Two-level model: red dashed bars. Corresponding photo-absorption profile, see text. Experimental absorbance profile for ZnO NC's of about 6 nm radius (Ref.~\cite{Alina1}): dot line.}
\label{fig3}
\end{figure}

In order to check the validity of the two-level approach and have a complete dipole response, we have performed a full RPA calculation with local exchange, following the method applied to metallic clusters by the present authors and references therein\cite{guet1, guet2}. The mean field potential that yields only radially node-less eigenfunctions in the ground state, has high excited states with one or more radial nodes, thus allowing for dipole transitions in addition to that singled out in the schematic model.  The full RPA calculation accounts for all-possible one-particle-one-hole excitations in the perturbing wave functions, thus including interactions between node-less occupied and one or more node unoccupied single-particle states.
Fig.~\ref{fig3} shows the RPA $f$ spectrum for systems of 8, 32, 50 and 128 electrons. There is a strong transition that lies at precisely the frequency obtained with the two-level model, as also shown in Fig.~\ref{fig2}.  This main transition carries about two-third of the total oscillator strength just as expected from the two-level model. This high concentration of strength at one frequency reveals a strong collectivity of the dipole resonance. However, it is an almost pure transition, governed by the particle-hole interaction between the the degenerate HOO and LUO, with practically no admixture of interactions with virtual states which necessarily would have one node radial wave function. The remaining oscillator strength is shared by transitions at higher frequencies, well separated from the main peak; these excitations involve particle-hole interaction matrix elements coupling node-less radial states with one-node    single-particle states and thus an overlap of almost orthogonal radial wave functions, which make them smaller by order of $r_0/R$ than particle-hole interaction matrix elements taken only over same node-less radial wave functions as in Eq.(~\ref{omega}). While nicely peaked at the theoretically predicted frequency, the experimental absorbance spectra are quite broad as shown in Fig.~\ref{fig3} for N$\sim$128, a feature which the present theory does not capture, unless we artificially fold the f-distribution with a Lorentzian of width equal to 0.4$\times$E. Apart from trivial broadening due to some experimental dispersion of numbers of carriers and NC radii, the observed width may have different contributions such as departure from sphericity, inhomogeneities of the interface, surface scattering, phonon coupling and finiteness of the band offset, all these contributions leading to fragmentation of the present model oscillator strength distribution. All these effects would require further theoretical studies that go beyond the present approach.
To conclude this Letter, we have worked out a theory that nicely predicts the strong dipole resonance observed in highly n-doped colloidal ZnO nanocrystals~\cite{Alina1}. In the limit of large radius, the resonance tends smoothly to the classical plasmon frequency of a charged shell of infinitesimal width. The repulsion between the quasi-electrons and the quantum containment of the latter within the NC well lead to an electron distribution resembling a spherical shell which radius departs from the NC radius with decreasing charge. This quantum "push-in" effect explains the observed blue shift. A purely independent-particle quantum model totally fails to yield the empirical data. It is essential to account for the induced polarization by the external field, in other words to take into account the residual interaction between particle-hole excitations. To conclude, highly-n-doped colloidal nano-crystals appear as a new paradigm to study the collective behavior of a strongly interacting N-electrons system confined in a quasi infinite spherical well.

L.G. and A.I. acknowledge Ministry of Education and Science of the Russian Federation (grant of Minobrnauka 16.9789.2017/BCh) for financial support.

\end{document}